\documentstyle[12pt]{article}

\newcommand{\eq}{\begin{equation}}
\newcommand{\en}{\end{equation}}
\newcommand{\eqa}{\begin{eqnarray}}
\newcommand{\ena}{\end{eqnarray}}
\newcommand{\eqan}{\begin{eqnarray*}}
\newcommand{\enan}{\end{eqnarray*}}

\newcommand{\lbl}{\label}

\newcommand{\sect}[1]{\setcounter{equation}{0}\section{#1}}

\newcommand{\NP}[1]{Nucl. Phys.\ {\bf #1}\ }
\newcommand{\PL}[1]{Phys. Lett.\ {\bf #1}\ }

\newcommand{\CMP}[1]{Comm. Math. Phys.\ {\bf #1}\ }
\newcommand{\PR}[1]{Phys. Rev\ {\bf #1}\ }
\newcommand{\PRE}[1]{Phys. Rep.\ {\bf #1}\ }

\newcommand{\JPL}[1]{J. Phys. (Paris)\ {\bf #1}\ }
\newcommand{\MPL}[1]{Mod. Phys. Lett.\ {\bf #1}\ }

\def\sqr#1#2{{\vcenter{\hrule height.#2pt
     \hbox{\vrule width.#2pt height#1pt \kern#1pt
        \vrule width.#2pt}
     \hrule height.#2pt}}}

\def\thinspace{\kern .16667em}
\def\Dir{\nabla\kern-7.8pt\Big{/}}

\def\reali{{\hbox{l\kern-.5mm R}}}
\def\naturali{{\hbox{l\kern-.5mm N}}}
\def\interi{{\hbox{Z\kern-1.5mm Z}}}
\def\complessi{{\bf C}}
\def\unity{{\hbox{\s@ 1\kern-.8mm l}}}
\def\uno{{\hbox{ 1\kern-.8mm l}}}

\def\part{\partial}

\def\aa{\alpha}
\def\bb{\beta}

\def\dd{\delta}
\def\DD{\Delta}
\def\ee{\epsilon}

\def\ff{\phi}
\def\FF{\Phi}
\def\vf{\varphi}
\def\gg{\gamma}
\def\GG{\Gamma}

\def\ll{\lambda}

\def\OO{\Omega}

\def\OB{\bar\Omega}
\def\pp{\psi}
\def\PP{\Psi}
\def\pb{\bar\psi}
\def\rr{\rho}

\def\ss{\sigma}

\def\tt{\theta}

\def\jb{{\bar j}}
\def\kb{{\bar k}}
\def\mb{{\bar m}}

\begin{document}
\begin{titlepage}
\begin{flushright}
DFTT 29/92\\
July 1992\\
Preliminary Version\\
hep-th/9207021
\end{flushright}
\vspace*{0.5cm}
\begin{center}
{\bf
\begin{Large}
{\bf
POLYMERS AND TOPOLOGICAL FIELD THEORY :
A 2 LOOP COMPUTATION\\}
\end{Large}
}
\vspace*{1.5cm}
         {\large I. Pesando}\footnote{E-mail I\_PESANDO@TO.INFN.IT ,
                                             39163::I\_PESANDO}
         \\[.5cm]
         Dipartimento di Fisica Teorica dell'Universit\`{a} di Torino
            \footnote{Address after 1 October 1992: Niels Bohr Institute,
                      Blegdamsvej 17, 2100 Copenhagen, Denmark}\\
         Istituto Nazionale di Fisica Nucleare, Sezione di Torino\\
         via P.Giuria 1, I-10125 Torino, Italy
\end{center}
\vspace*{0.7cm}
\begin{abstract}
{\large
Within the Quantum Action Principle framework we show the perturbative
renormalizability of previously proposed topological lagrangian \`a la
Witten-Fujikawa describing polymers, then we perform a 2 loop
computation.
The theory turns out to have the same predictive power of De Gennes
theory, even though its running coupling constants exhibit a very
peculiar behaviour.
Moreover we argue that the theory presents two phases , a
topological and a non topological one.
}
\end{abstract}
\vfill
\end{titlepage}

\setcounter{footnote}{0}
\section*{Introduction.}
In a previous work \cite{PC} we proposed a topological theory \`a la
Fujikawa-Witten \cite{Fu,Wi,Bi} describing the self--avoiding walks
(hereafter SAW), i.e. the polymers in the De Gennes model \cite{DG,DCl}.

The aim of this approach was the exact computation of the critical
exponents of SAW; this would have been achieved by an exact computation
of the theory $\bb$-function(s) relying on the "topologicity" of the
theory.

In this article we want to prove that the theory we proposed is actually
perturbatively renormalizable and to perform the calculation of the
interesting quantities up to the second loop.
The explicit computation reveals that , in spite of the topologicity of
the theory, the hope for an exact computation of $\bb$-function is not
fulfilled, nonetheless the theory has some interesting features such as
the doubling of the coupling constants, which exhibit a very peculiar
behaviour under the RG flow, and the possibility for a spontaneous
breaking of the topological phase.

The article is divided as follows:
in section one we review the relation among our model and those of De
Gennes (\cite{DG,DCl}) and of Parisi-Sourlas-Mc Kane (\cite{MK,PS});
in section two we prove of the
perturbative renormalizability in the formalism of Quantum Action
Principle {\cite{QAP}-\cite{QAP1});in section three we explain the two loop
computation in the framework of background field method (\cite{bck});
 in section four we discuss the renormalization group flow,
finally we draw our conclusions.

\def\dag{\dagger}
\def\FP{\Phi\Pi}

\sect{The topological theory of polymers.}
%In a previous work \cite{PC} we proposed a topological theory
%describing the SAW , i.e. the polymers;
We start discussing briefly the relation between our model and the MPS
one.
The renormalization requires a
slight generalization of the previous lagrangian \cite{PC} that can
easily understood as the necessity of including all the BRST invariant
terms with the same dimension.
The lagrangian is now given by:
$${\cal L}=
\rr b^\dag b
+i b^\dag   (-\DD+m^2)\ff
+i \ff^\dag (-\DD+m^2)b
-i \xi^\dag (-\DD+m^2)\eta
+i \eta^\dag(-\DD+m^2)\xi
$$
\eq
+\ll\left(~ib^{\dagger}\ff-i~\ff^{\dagger}b
      -i~\xi^{\dagger}\eta-i~\eta^{\dagger}\xi\right)^2
+\nu\left(~b^{\dagger}\ff+\ff^{\dagger}b
      -\xi^{\dagger}\eta+\eta^{\dagger}\xi\right)^2
\lbl{laga}
\en
where $b=(b_1,\dots, b_N)^T,\ff=(\ff_1,\dots,\ff_N)^T$ are two vectors
of N complex Lorentz scalars (N being an arbitrary natural number),
$\xi=(\xi_1,\dots,\xi_N)^T$, $\eta=(\eta_1,\dots,\eta_N)^T$ are two
vectors of N complex Lorentz scalar ghosts, and $\rr$,
$\ll$ and $\nu$ are arbitrary positive numbers satisfying the condition
$\ll > \nu$.
The dimension and the BRST charge of the fields is given in the
following table:

\centerline{
\begin{tabular}{|c|c|c|c|c|} \hline
           & $\ff$  &  $b$   & $\xi$  & $\eta$ \\ \hline
dim        &   0    &  2     &    1   &    1   \\ \hline
$\FP$      &    0   &  0     &   -1   &    1   \\ \hline
\end{tabular}
}
\vskip.5cm
\noindent
and discussed in the next section.
The two adimensional coupling constants $\ll$ and $\nu$ are related to
the $O(n\rightarrow 0)$ coupling constant $g$ by
\eq
g=\ll-\nu
\lbl{rel}
\en
this implies that $\ll=\nu$ is a complicate way of describing a free
theory.
The proof of this relation and of the formal equivalence between De
Gennes theory and the present one is based upon the equality of the two
points Green function of the two theories; this can easily be achieved
by rewriting (\ref{laga}) using two auxiliary fields $\aa,\bb$ as
\eq
{\cal L}=
\rr b^\dag b
+i b^\dag O \ff    +i \ff^\dag O^\dag b
-i \xi^\dag O \eta +i \eta^\dag O^\dag \xi
+\aa^2+\bb^2
\lbl{prova}
\en
where
$$
O=   (-\DD+m^2+2i\sqrt{\ll}\aa+2\sqrt{\nu}\bb)
$$
and by using the McKane-Parisi-Sourlas trick \cite{MK,PS} on De Gennes
theory \cite{DG,DCl} along with (\ref{rel}), in such a way that De
Gennes theory can be rewritten as:
$${\cal L}_{O(n\rightarrow0)}=
\ff^\dag   O \ff
+\psi^\dag O \psi
+\aa^2+\bb^2
$$
We want to stress that this equivalence is true only if $\ll,\nu >0$
because otherwise, after integrating over $b$ in (\ref{prova}), the
remaining effective action would not be bounded from below.

\noindent
For computational purpose it is better to rewrite (\ref{laga}) as
$${\cal L}=
\rr~ \vf^\dag U\vf
+i \vf^\dag  Y(-\DD+m^2)\vf
- \psi^\dag X(-\DD+m^2)\psi
$$
\eq
+\ll(\vf^\dag X\vf -i \psi^\dag Y\psi)^2
+\nu(\vf^\dag Y\vf +i \psi^\dag X\psi)^2
\lbl{lag}
\en
where we introduced the following matrices
$$
X=\left(\begin{array}{cc}
         0 & i\uno_N \\
        -i\uno_N & 0
        \end{array}
   \right)=\ss_1\otimes\uno_N
{}~~~~
Y=\left(\begin{array}{cc}
         0 & \uno_N\\
         \uno_N & 0
        \end{array}
   \right)=\ss_2\otimes\uno_N
$$$$
U=\left(\begin{array}{cc}
         \uno_N & 0\\
         0      & 0
        \end{array}
   \right)
$$
and we defined
$$
\vf=\left( \begin{array}{c} b \\ \ff \end{array} \right )
{}~~~~
\psi=\left( \begin{array}{c} \xi \\ \eta \end{array} \right )
$$
Notice the symmetry in the four-field terms
$$ (\ll,X,Y)\leftrightarrow (\nu,Y,-X)$$
that turns out to be useful in performing the actual computation.

\def\L{{\cal L}}
\def\W{{\cal W}}
\def\S{{\cal S}}
\def\pam{\partial^\mu}
\def\pan{\partial^\nu}

\def\D{{\cal D}}
\def\F{{\cal F}}
\sect{Symmetries and renormalization}
In the previous section we gave the topological lagrangian along with
the dimension of the fields and their BRST charge, without motivating
these choices and the terminology.

Now we proceed in explaining the field dimensions, they are deduced
looking at the explicit form of the free theory propagators:
\eqa
<\ff^*_j (x)\ff_k (0)>=\dd_{jk}\int~d^4k~e^{ikx}{\rr\over(k^2+m^2)^2}
\nonumber\\
<\ff^*_j (x) b_k (0)>=\dd_{jk}\int~d^4k~e^{ikx}{-i\over(k^2+m^2)}
\nonumber\\
<\xi^*_j (x) \eta_k (0)>=\dd_{jk}\int~d^4k~e^{ikx}{-i\over(k^2+m^2)}
\ena
One could wonder about the dimensionality of the parameter $\rr$ and,
consequently, of the other fields, but what justifies setting to zero
the dimensionality of $\rr$ is just the power counting in which the
propagator $\ff\ff$ has dimension -4 (in unit of mass).
Just because of this noncanonical dimension of the $\ff\ff$ propagator
% it is worth redoing
we performed the usual power counting both with $\rr\ne0$ and $\rr=0$.
It is easy to show that the critical dimension is four (as it should be
to reproduce De Gennes theory) and that if we
indicate with $E_\ff,E_b,E_\xi,E_\eta$ the number of external legs
 of the fields $\ff,b,\xi,\eta$ in a truncated diagram,
at the critical dimensionality $D_{\mbox{cr}}=4$,
there are only the following superficially divergent diagrams:
%\centerline{
\eq
\begin{tabular}{|c|c|c|c|}\hline
$E_b$ &$E_\ff$ &$E_\xi=E_\eta$& D \\ \hline
 0     &  0     &    0         & 4 \\ \hline
 1     &  1     &    0         & 2 \\ \hline
 0     &  0     &    1         & 2 \\ \hline
 1     &  1     &    1         & 0 \\ \hline
 2     &  0,2   &    0         & 0 \\ \hline
 0     &  0     &    2         & 0 \\ \hline
\end{tabular}
\lbl{tabella}
\en
%}

We are ready to discuss both the symmetries and
the broken symmetries of the action:
\begin{enumerate}
\item the discrete symmetry:
\eq
      \ff\rightarrow\ff^*,b\rightarrow b^*,\xi\rightarrow\xi^*,
      \eta\rightarrow\eta^*
\en
      that is responsible for the non appearance
      in the lagrangian of a term
      $$
      \left(~ib^{\dagger}\ff-i~\ff^{\dagger}b
      -i~\xi^{\dagger}\eta-i~\eta^{\dagger}\xi\right)
      \left(~b^{\dagger}\ff+\ff^{\dagger}b
      -\xi^{\dagger}\eta+\eta^{\dagger}\xi\right)
      $$
       which is odd under such a transformation;
      this symmetry allows to construct the action
      only from the real part of functionals
      (possibly with complex coefficients)
      and from  even power of the imaginary part of functionals
      (possibly with complex coefficients);
\item $GL(N,\complessi)_{\mbox{bos}}$ broken to $U(N,\reali)$
      by the term proportional to $b^\dag b$:
\eqa
      \dd^{b}(\tt^\aa)\ff &=&i~G_\aa\ff~\tt^\aa
      ~~~~
      \dd^{b}(\tt^\aa)\ff^\dag =-i~\ff^\dag G^\dag_\aa\ff~\tt^{\aa *}
 \nonumber\\
      \dd^{b}(\tt^\aa) b^\dag &=& -i~b^\dag G_\aa~\tt^\aa
      ~~~~
      \dd^{b}(\tt^\aa) b = i~G^\dag_\aa b~\tt^{\aa *}
\ena
      where $G_\aa$ are the generators of $gl(N,\complessi)$
      and $\tt_{\aa}$ is the complex parameter of the transformation.

      \noindent
      We use the following explicit representation for the generators
      $$
      (G_\aa)_{pq}\equiv (G_{(ab)})_{pq}=-i\dd_{ap}\dd_{bq}
      ~~~~
      ( G_{(ab)}^\dag=-G_{(ba)} )
      $$
      \noindent
      Considering $gl(N)$ over the complex field allows to vary
      independently $(\ff,b^*)$ from $(\ff^*,b)$, in fact we can build
      the following generators of the decomplexified algebra:
\eqa
      \dd_{\aa}^{(+)}={1\over2}(\dd_{\aa}(\tt_{\aa}=1)
                        -i(\dd_{\aa}(\tt_{\aa}=i) )
\nonumber\\
      \dd_{\aa}^{(-)}={1\over2}(\dd_{\aa}(\tt_{\aa}=1)
                        +i(\dd_{\aa}(\tt_{\aa}=i) )
\ena
      whose action on the bosonic fields is given by
\eqa
      \dd_{(ab)}^{(+)}\ff_p=\dd_{ap}\ff_b,
      ~~~~
      \dd_{(ab)}^{(+)}b^*_p=-\dd_{bp} b^*_a,
      ~~~~
      \dd_{(ab)}^{(+)}\ff^*_p=\dd_{(ab)}^{(+)}b_p=0
 \nonumber\\
      \dd_{(ab)}^{(-)}\ff^*_p=\dd_{ap}\ff^*_b,
      ~~~~
      \dd_{(ab)}^{(-)}b_p=-\dd_{bp} b_a,
      ~~~~
      \dd_{(ab)}^{(-)}\ff_p=\dd_{(ab)}^{(-)}b^*_p=0
\ena

      This symmetry is however broken ,
      in fact we find immediately the breaking under the transformation
      by a complex parameter $\tt^\aa$:
      $$ \dd {\cal L}=
        i~\rr~b^\dag(G^\dag_\aa-G_\aa)b ~Re \tt^\aa
        +\rr~b^\dag(G^\dag_\aa+G_\aa)b ~Im \tt^\aa
      $$
      the breaking vanishes when restricting the
      symmetry to the $u(N,\reali)$ generated by $T_A$ whose explicit
      representation is given by:
      $$T_{(aa)}=iG_{(aa)}$$
      $$T_{1(ab)}=G_{(ab)}+G_{(ab)}^\dag
      ~~~~
      T_{2(ab)}=i(G_{(ab)}-G_{(ab)}^\dag)
      ~~~~
      \mbox{ with } a<b
      $$.
\item $GL(N,\complessi)_{\mbox{ferm}}$ with complex parameter $\tt_\aa$:
\eqa
      \dd^{f}(\tt^\aa)\eta &=& i~G_\aa\eta ~\tt^\aa
      ~~~~
      \dd^{f}(\tt^\aa)\eta^\dag =-i\eta^\dag G_\aa^\dag~\tt^{\aa *}
\nonumber\\
      \dd^{f}(\tt^\aa) \xi^\dag &=& -i~ \xi^\dag G_\aa~\tt^\aa
      ~~~~
      \dd^{f}(\tt^\aa) \xi = i~ G_\aa^\dag \xi~\tt^{\aa *}
\ena
      As in the previous case of the broken bosonic $GL(N,\complessi)$,
      it is possible to build $\dd^{f(\pm)}$ and vary independently
      the couple $(\eta,\xi^*)$ from $(\eta^*,\xi)$.
\item BRST-like transformations with complex parameter $\tt^\aa$:
\eqa
      \hat\dd(\tt^\aa) \ff=iG_\aa\eta~\tt^\aa ,
      ~~~~
      \hat\dd(\tt^\aa)\eta=0
\nonumber\\
      \hat\dd(\tt^\aa)  \xi^\dag=i b^\dag G_\aa~\tt^\aa ,
      ~~~~
      \hat\dd(\tt^\aa)  b=0
\ena
      In particular the generator of the "canonical" BRST is
      $s=\sum_{a=1}^N \hat\dd_{(aa)}$, in such a way we can rewrite the
      lagrangian (\ref{laga}) as
      $${\cal L}=s[\rr~b^\dag\xi
        +i \xi^\dag   (-\DD+m^2)\ff
        +i \ff^\dag (-\DD+m^2)\xi
      $$$$
     +\ll\left(~ib^{\dagger}\ff-i~\ff^{\dagger}b
      -i~\xi^{\dagger}\eta-i~\eta^{\dagger}\xi\right)
     \left(~i\xi^\dag\ff-i\ff^\dag\xi\right)
      $$$$
     +\nu\left(~b^{\dagger}\ff+\ff^{\dagger}b
      -\xi^{\dagger}\eta+\eta^{\dagger}\xi\right)
     \left(~\xi^\dag\ff+\ff^\dag\xi\right)]
      $$
    Notice that the explicit form and existence of $s$ justifies the
     dimensions and the charges of the fields  we gave.
     Exactly as before, we can build $\hat\dd^{(\pm)}$ and vary
    independently $(\ff,\xi^*)$ from $(\ff^*,\xi)$.

\item antiBRST-like transformations broken by $b^\dag b$:
\eqa
      \widehat{\bar\dd_\aa}\eta=iG^\dag_\aa\ff,
      ~~~~\widehat{\bar\dd_\aa}\ff=0 \nonumber\\
      \widehat{\bar\dd_\aa} b^\dag=-i \xi^\dag G_\aa^\dag,
      ~~~~\widehat{\bar\dd_\aa} \xi=0
\ena
      The breaking is :
      $$ \widehat{\bar\dd}(\tt^\aa){\cal L}=
       i~\rr~(b^\dag G^\dag_\aa\xi -\xi^\dag G_\aa b)~Re(\tt^\aa)
       +~\rr~(b^\dag G^\dag_\aa\xi +\xi^\dag G_\aa b)~Im(\tt^\aa)
       $$
\end{enumerate}

Since the broken symmetries are broken by a term of dimension four, it
would be very difficult to keep them under control, so we prefer to give
up these symmetries and to consider only the unbroken ones.

In order to implement the Ward-Takahashi identities (WTI) we
introduce the following functional operators:
\begin{enumerate}
\item
\eq
      u(N,\reali)\Longrightarrow
      \W^b_A=\int~i(T_A\ff)_p{\dd\over\dd\ff_p}
      -i(b^\dag T_A)_p{\dd\over\dd b^*_p}+(c.c.)
\en
\item
\eq
      gl(N,\complessi)\Longrightarrow
      \left\{
      \begin{array}{l}
      \W^{f(+)}_\aa=\int~i(G_\aa\eta)_p{\vec{\dd}\over\dd\eta_p}
      -i(\xi^\dag G_\aa)_p{\vec{\dd}\over\dd \xi^*_p}
      \\
      \W^{f(-)}_\aa=\int~i(G^\dag_\aa \xi)_p{\vec{\dd}\over\dd \xi_p}
      -i(\eta^\dag G^\dag_\aa)_p{\vec{\dd}\over\dd \eta^*_p}
      \end{array}
      \right.
\en
\item
\eq
      \mbox{BRST-like} \Longrightarrow
      \left\{
      \begin{array}{l}
      \S^{(+)}_\aa=\int~i(G_\aa\eta)_p{\dd\over\dd\ff_p}
      +i(b^\dag G_\aa)_p{\vec{\dd}\over\dd \xi^*_p}
      \\
      \S^{(-)}_\aa=\int~i(G^\dag_\aa b)_p{\vec{\dd}\over\dd\xi_p}
      +i(\eta^\dag G^\dag_\aa)_p{\dd\over\dd \ff^*_p}
      \end{array}
      \right.
\en
\end{enumerate}

\noindent
Notice that sine the fields are doublets under the BRST it is not
necessary to introduce external sources for different kind of
BRST multiplets.

\noindent
All the symmetries are contained in the following WTI:
\eqa
{[}\W^b_A,\W^b_B ]\GG
   &=&f_{AB}^{~~~C} \W^b_C \GG
\label{wtglbos}
\\
{[}\W^b_A,\W^{f(\pm)}_\bb ] \GG
   &=& 0
\\
{[}\W^b_A, \S^{(\pm)}_\bb] \GG
   &=& g_{A\bb}^{~~~\gg(\pm)} \S^{(\pm)}_\gg\GG
\\
 {[} \W ^{f(\pm)}_\aa,\W ^{f(\pm)}_\bb {]}\GG
   &=&  f_{\aa\bb}^{~~~\gg}\W^{f(\pm)}_\gg\GG
\\
 {[} \W ^{f(\pm)}_\aa,\W ^{f(\mp)}_\bb {]}\GG
   &=& 0
\\
 {[} \W ^{f(\pm)}_\aa,\S^{(\pm)}_\bb {]}\GG
   &=& g_{\aa |\bb}^{~~~\gg(\pm)}\W^{f(\pm)}_\gg\GG
\\
 {[} \W ^{f(\pm)}_\aa,\S^{(\mp)}_\bb {]}\GG
   &=&
\\
 {[} \S^{(\pm)}_\aa,\S^{(\pm)}_\bb ]_+\GG
&=&
 {[} \S^{(\pm)}_\aa,\S^{(\mp)}_\bb ]_+\GG
 =0
\lbl{wtss}
\ena
where
\eqa
{[}T_A,T_B]&=&i~f_{AB}^{~~~C}T_C
\nonumber\\
{[}G_\aa,G_\bb]&=&i~f_{\aa\bb}^{~~~\gg}G_\gg
\nonumber\\
T_AG_\aa &=&
i~g_{A\aa}^{~~~\bb(+)}G_\bb=
i~g_{A\aa}^{~~~\bb(-)*}G_\bb
\nonumber\\
G_\aa G_\bb &=&
-i~g_{\aa |\bb}^{~~~\gg(+)}G_\gg=
i~g_{\aa |\bb}^{~~~\gg(-)*}G_\gg
\nonumber
\ena
It can be shown (with a big amount of algebra) that these symmetries
are not anomalous.

After the discussion of the WTI we can discuss the stability
of the classical action $\GG_{\mbox{cl}}$.
This amount to impose the conditions (\ref{wtglbos}-\ref{wtss}) to the
perturbed action $\GG '=\GG_{\mbox{cl}}+\DD$.

The $U(1)^N_f$ symmetry ( $\W^f_{(aa)}\DD=0$) implies that every term of
$\DD$ has to be built using an equal number of conjugate ghost fields
 and  ghost fields\footnote{
This assertion is intuitively obvious, however a rigorous proof is
based on the observation that
$\W^f_{(aa)}=N(\eta_a)-N(\eta^*_a)+N(\xi^*_a)-N(\xi_a)$
where $N(\eta_a)=\int \eta_a{\vec{\dd}\over\dd\eta_a}$ can be interpreted
as an occupation number operator.},
in the mean time the ghost charge implies that
every term should contain an equal number of $\xi$ and $\eta$.
Taking also in account the discrete symmetry, the dimension of the fields
 and the fact that we are looking
at an integrated functional, the explicit most general form of $\DD$ is
\eqa
\DD&=&\int
\xi^*_a\eta_b ~f_{ab}[\ff,\ff^*,b,b^*]
+\xi_a\eta^*_b ~f_{ab}[\ff^*,\ff,b^*,b]
\nonumber\\
&&
+\pam\xi^*_a\eta_b ~g_{\mu ab}[\ff,\ff^*]
+\pam\xi_a\eta^*_b ~g_{\mu ab}[\ff^*,\ff]
\nonumber\\
&&
+\pam\xi^*_a\pan\eta_b ~h_{\mu\nu ab}(\ff,\ff^*)
+\pam\xi_a\pan\eta^*_b ~h_{\mu\nu ab}(\ff^*,\ff)
\nonumber\\
&&
+\xi^*_a\xi^*_b\eta_c\eta_d ~l_{1~ab|cd}(\ff,\ff^*)
+\xi_a\xi_b\eta^*_c\eta^*_d ~l_{1~ab|cd}(\ff^*,\ff)
\nonumber\\
&&
+\xi^*_a\xi_b\eta^*_c\eta_d ~l_{2~a|b|c|d}(\ff,\ff^*)
+n[\ff,\ff^*,b,b^*]
\ena
{}From Lorentz invariance and applying $\W^{f(\pm)}_{(ab)}~~(a\ne b)$
to this expression, it reduces to:
\eqa
\DD&=&\int
\xi^\dag   \eta ~f[\ff,\ff^*,b,b^*]
-\eta^\dag \xi ~f[\ff^*,\ff,b^*,b]
\nonumber\\
&&
+\pam\xi^\dag\eta ~
(\part_\mu\ff g^{(1)}(\ff,\ff^*)+\part_\mu\ff^* g^{(2)}(\ff,\ff^*))
\nonumber\\
&&
-\eta^\dag\pam\xi ~
(\part_\mu\ff^* g^{(1)}(\ff^*,\ff)+\part_\mu\ff g^{(2)}(\ff^*,\ff))
\nonumber\\
&&
+\pam\xi^\dag\part_\mu\eta ~h(\ff,\ff^*)
-\pam\eta^\dag\part_\mu\xi ~h(\ff^*,\ff)
+\xi^\dag\eta\eta^\dag\xi ~l_{2}(\ff,\ff^*)
\nonumber\\
&&
+(\xi^\dag\eta)^2 ~l_{1}(\ff,\ff^*)
+(\eta^\dag\xi)^2 ~l_{1}(\ff^*,\ff)
+n[\ff,\ff^*,b,b^*]
\ena
{}From $U(1)^N_b\otimes U(1)^N_f$ and
$\S^{(\pm)}\DD|_{\part^0}=0$ in the sector without derivatives, we get
$l_1,l_2$ constants and:
$$
\left. \DD\right|_{\part^0}=
\int -{1\over4}(2l_1+l_2)
\left(b^{\dagger}\ff-\ff^{\dagger}b
      -\xi^{\dagger}\eta-\eta^{\dagger}\xi\right)^2
$$$$
+ {1\over4}(2l_1-l_2)
\left(~b^{\dagger}\ff+\ff^{\dagger}b
      -\xi^{\dagger}\eta+\eta^{\dagger}\xi\right)^2
$$
\eq
+n_0 b^\dag b
-f_0 (b^\dag\ff+\ff^\dag b -\xi^\dag\eta+\eta^\dag\xi)
\en
where $f_0$ and $n_0$ are constants.
Examining the sector with two derivatives, it is easy to realize
that terms proportional to $m^2$, i.e. with the structure
$m^2\part^2\ff$, are absent;
then from $\S^{(\pm)}\DD|_{b\part^2}=\S^{(\pm)}\DD|_{b^*\part^2}=0$, it
is not difficult to prove that
$$
\left. \DD\right|_{\part^2}=
n~\int
b^\dag\DD\ff
+ \ff^\dag \DD b
- \xi^\dag \DD\eta
+ \eta^\dag \DD\xi
$$
Finally we can set immediately to zero the four derivatives part
of $\DD$ because it is impossible to have diagrams with $p^4$ behaviour
(\ref{tabella}).

\def\fc{\phi^\dagger}
\def\FC{\Phi^\dagger}
\def\PC{\Psi^\dagger}
\def\pc{\psi^\dagger}

\sect{The quantum corrections.}

In the following we will use the background field method (\cite{bck}).
To this aim we split the fields as follows:
\eq
\vf\rightarrow \FF+\vf_{\mbox{quant}}
{}~~~~
\pp\rightarrow \PP+\pp_{\mbox{quant}}
\lbl{redef}
\en
where $\FF,\PP$ are the classical background fields.
Performing this splitting  (and dropping the specification quant)
the lagrangian (\ref{lag}) becomes:
$${\cal L}_{\mbox{quant}}=
+i \vf^\dag  Y(-\DD)\vf
- \psi^\dag X(-\DD)\psi
$$$$
+\vf^T M_1 \vf
+\vf^\dag M_2 \vf
+\vf^\dag M_3 \vf^*
+\pp^T N_1 \pp
+\pp^\dag N_2 \pp
+\pp^\dag N_3 \pp^*
$$$$
+\vf^T \OB_1 \pp
+\vf^\dag \OB_2 \pp
+\pp^\dag \OO_1 \vf
+\pp^\dag \OO_2 \vf^*
$$$$
+A_{2ij\kb}\vf_i\vf_j\vf^*_\kb
+A_{3i\jb\kb}\vf_i\vf^*_\jb\vf^*_\kb
+{\cal A}_{2ij\kb}\pp_i\pp_j\pp^*_\kb
+{\cal A}_{3i\jb\kb}\pp_i\pp^*_\jb\pp^*_\kb
$$$$
+{\cal B}_{3i\jb k}\vf_i\vf^*_\jb\pp_k
+{\cal B}_{4i\jb\kb}\vf_i\vf^*_\jb\pp^*_\kb
+B_{3i\jb k}\pp_i\pp^*_\jb\vf_k
+B_{4i\jb\kb}\pp_i\pp^*_\jb\vf^*_\kb
$$
\eq
+C_{ij\kb\mb}\vf_i\vf_j\vf^*_\kb\vf^*_\mb
+D_{ij\kb\mb}\pp_i\pp_j\pp^*_\kb\pp^*_\mb
+{\cal E}_{i\jb k \mb}\vf_i\vf^*_\jb\pp_k\pp^*_\mb
\lbl{lagbk}
\en
where all the coefficients can be obtained easily from (\ref{lag}) using
(\ref{redef});
explicitly  we get
%for instance:
\eqa
\vf^T M_1 \vf&=&
     \ll \FC X \vf ~\FC X \vf
    +\nu \FC Y \vf ~\FC Y \vf
%+( (\ll,X,Y)\leftrightarrow (\nu,Y,-X))
\nonumber\\
M_2&=&a~U
   +2\ll(\FC X\FF-i\PC Y\PP) X
\nonumber\\
&&
   +[i~m^2+2\nu(\FC Y\FF+i\PC X\PP)] Y
\nonumber\\
  && +2\ll X\FF \FC X
   +2\nu Y\FF \FC Y
\nonumber\\
\vf^\dag M_3 \vf^*&=&
    \ll \fc X \FF ~\fc X \FF
   +\nu \fc Y \FF ~\fc Y \FF
\nonumber\\
\pp^T N_1 \pp&=&
    -\ll \PC Y \pp ~\PC Y \pp
    -\nu \PC X \pp ~\PC X \pp
\nonumber\\
N_2&=&
   -2i\ll(\FC X\FF-i\PC Y\PP) Y
\nonumber\\
  && +[-m^2+2i\nu(\FC Y\FF+i\PC X\PP)] X
\nonumber\\
  && -2\ll Y\PP \PC Y
   -2\nu X\PP \PC X
\nonumber\\
\pp^\dag N_3 \pp^*&=&
   -\ll \pc Y \PP ~\pc Y \PP
   -\nu \pc X \PP ~\pc X \PP
\nonumber\\
\vf^T \OB_1 \pp&=&
    2i\ll \FC X \vf ~\PC Y \pp
   -2i\nu \FC Y \vf ~\PC X \pp
\nonumber\\
\vf^\dag \OB_2 \pp&=&
    2i\ll \fc X \FF ~\PC Y \pp
   -2i\nu \fc Y \FF ~\PC X \pp
\nonumber\\
\pp^\dag \OO_1 \vf&=&
    2i\ll \FC X \vf ~\pc Y \PP
   -2i\nu \FC Y \vf ~\pc X \PP
\nonumber\\
\pp^\dag \OO_2 \vf^*&=&
    2i\ll \fc X \FF ~\pc Y \PP
   -2i\nu \fc Y \FF ~\pc X \PP
\nonumber\\
A_{2ij\kb}\vf_i\vf_j\vf^*_\kb&=&
    2\ll \fc X \vf ~\FC X \vf
   +2\nu \fc Y \vf ~\FC Y \vf
\nonumber\\
A_{3i\jb\kb}\vf_i\vf^*_\jb\vf^*_\kb&=&
    2\ll \fc X \vf ~\fc X \FF
   +2\nu \fc Y \vf ~\fc Y \FF
\nonumber\\
{\cal A}_{2ij\kb}\pp_i\pp_j\pp^*_\kb&=&
   -2\ll \pc Y \pp ~\PC Y \pp
   -2\nu \pc X \pp ~\PC X \pp
\nonumber\\
{\cal A}_{3i\jb\kb}\pp_i\pp^*_\jb\pp^*_\kb&=&
   -2\ll \pc Y \pp ~\pc Y \PP
   -2\nu \pc X \pp ~\pc X \PP
\nonumber\\
{\cal B}_{3i\jb k}\vf_i\vf^*_\jb\pp_k&=&
   -2i\ll \fc X \vf ~\PC Y \pp
   +2i\nu \fc Y \vf ~\PC X \pp
\nonumber\\
{\cal B}_{4i\jb\kb}\vf_i\vf^*_\jb\pp^*_\kb&=&
   -2i\ll \fc X \vf ~\pc Y \PP
   +2i\nu \fc Y \vf ~\pc X \PP
\nonumber\\
B_{3i\jb k}\pp_i\pp^*_\jb\vf_k&=&
   -2i\ll \FC X \vf ~\pc Y \pp
   +2i\nu \FC Y \vf ~\pc X \pp
\nonumber\\
B_{4i\jb\kb}\pp_i\pp^*_\jb\vf^*_\kb&=&
   -2i\ll \fc X \FF ~\pc Y \pp
   +2i\nu \fc Y \FF ~\pc X \pp
\nonumber\\
C_{ij\kb\mb}\vf_i\vf_j\vf^*_\kb\vf^*_\mb&=&
   \ll \fc X \vf ~\fc X \vf
   +\nu \fc Y \vf ~\fc Y \vf
\nonumber\\
D_{ij\kb\mb}\pp_i\pp_j\pp^*_\kb\pp^*_\mb&=&
   -\ll \pc Y \pp ~\pc Y \pp
   -\nu \pc X \pp ~\pc X \pp
\nonumber\\
{\cal E}_{i\jb k \mb}\vf_i\vf^*_\jb\pp_k\pp^*_\mb&=&
   -2i\ll \fc X \vf ~\pc Y \pp
   +2i\nu \fc Y \vf ~\pc X \pp
\ena
\noindent
We performed the computation of Feynman graphs
in $D=4-2\ee$ and within the $\overline{\mbox{MS}}$ scheme.
In fig.s 1,2,3,4 \footnote{ Dashed lines are $\pb \pp$ propagators,
continuous lines are $\vf^* \vf$ propagators.}
two representatives for each of
four different kinds of graphs involved in the computation are given.
Here we want only to point out that the diagrams like those of
of Fig.~2 give $Z_\vf$, those similar to those of Fig.~3
(i.e. those containing at least either one factor $M_2$
 or one $N_2$) generate and cancel the overlapping divergences
proportional to currents similar to
$\FC (x) X \FF (x)~\FC (y) X \FF (y)$
while the graphs (like those ) of Fig. 4 cancel the overlapping
divergences containing currents of the kind
$\FC (x) X \FF (y)~\FC (y) X \FF (x)$.

\noindent
The one loop computation (graphs of Fig.1 ) yields :
\eqa
\dd_{(1)}\ll&=&{1\over (4\pi)^2\ee}4\ll(\ll-3\nu)
\nonumber\\
\dd_{(1)}\nu&=&-{1\over (4\pi)^2\ee}4(\ll^2-\ll\nu+2\nu^2)
\nonumber\\
\dd_{(1)}m^2&=&{1\over (4\pi)^2\ee}2m^2(\ll-\nu)
\nonumber\\
\dd_{(1)}\rr &=&-{1\over (4\pi)^2\ee}2\rr (\ll+\nu)
\lbl{1loop}
\ena

\noindent
The two loops computation yields:
\eqa
\dd_{(2)}\ll&=&{1\over (4\pi)^4}
               \left[{4\ll\over\ee^2}(10\ll^2-24\ll\nu+30\nu^2)
                    +{4\ll\over\ee}(-5\ll^2+18\ll\nu-21\nu^2)\right]
\nonumber\\
\dd_{(2)}\nu&=&-{1\over (4\pi)^4}
               \left[{4\over\ee^2}(6\ll^3-24\ll^2\nu+18\ll\nu^2-16\nu^3)
                    +{4\over\ee}(-6\ll^2+15\ll^2\nu-12\ll\nu^2+11\nu^3)
               \right]
\nonumber\\
\dd_{(2)}m^2&=&{1\over (4\pi)^4}m^2
               \left[{10\over\ee^2}-{6\over\ee}\right](\ll-\nu)^2
\nonumber\\
\dd_{(2)}\rr &=&-{1\over (4\pi)^4}2\rr
               \left[-{1\over\ee^2}(\ll^2+6\ll\nu+5\nu^2)
                    +{1\over\ee}(-\ll^2+2\ll\nu+3\nu^2)
               \right]
\nonumber\\
Z_\vf^{(2)}&=&
Z_b^{(2)}=Z_\xi^{(2)}=Z_\eta^{(2)}=
-{1\over (4\pi)^4\ee}(\ll-\nu)^2
\lbl{2loop}
\ena
where $Z_\vf$ is the wave function renormalization (
$\vf_{bare}=Z^{1/2}_\vf \vf_{ren}$).

\noindent
One could wonder why setting $Z_\vf=Z_b=Z_\xi=Z_\eta$ when $b$ and
$\vf$ have a different dimension; the answer lies in the fact that
renormalization fixes $Z_\rr Z_b$, $Z_b Z_\vf=Z_\xi Z_\eta$,
$Z_\ll (Z_b Z_\vf)^{1/2}$ and $Z_\nu (Z_b Z_\vf)^{1/2}$, while leaving
two free parameters ( $Z_b$ and $Z_\eta$, for instance).
This arbitrariness is however easily understood as the possibility of
redefining $b$ and $\eta$ inside the path integral; because of this
interpretation, this arbitrariness does not affect the physics.
Notice that there is also another natural choice for the free parameters:
$Z_\rr=1, Z_b=Z_\xi$, so that $\rr$ becomes a free constant
and not a coupling constant;
we want to stress that even in the delta gauge
($\rr=0$, \cite{Bi}) the theory does not become finite and the quantum
corrections to $\ll, \nu$ do not change.

\section{The RG flow.}
As it is easy to see $Z_\ff$, $\dd m^2$ and $\dd\ll-\dd\nu$ are
expressible as a function of $g=\ll-\nu$, in fact
from (\ref{1loop},\ref{2loop}) we get:
\eqa
Z_\ff&=&-{1\over (4\pi)^4\ee}g^2
\nonumber\\
\dd m^2&=&{1\over (4\pi)^2\ee}2m^2g
         +{1\over (4\pi)^4}m^2g^2
          \left({10\over\ee^2}-{6\over\ee}\right)
\nonumber\\
\dd g=\dd \ll -\dd \nu&=&
      {1\over (4\pi)^2\ee}8g^2
      +{1\over (4\pi)^4}g^3({64\over\ee^2}-{44\over\ee})
\ena
These are exactly the quantum corrections obtainable in the theory
$O(n\rightarrow 0)$ and this strongly suggests, even if it does
not prove, that the
perturbative expansion of $Z_\ff,\dd m^2$ and $\dd\ll-\dd\nu$ in our theory is
equal to that of the corresponding quantities in $O(n\rightarrow 0)$
theory \footnote{
Would we have chosen $Z_\rr '=1, Z_b '=Z_\xi '$, this would not have
been completely true, nevertheless what really matters, the physical
quantities, would have behaved exactly as $O(n\rightarrow 0)$ theory:
for instance, the two points function $<\ff^* b>$ depends only on
$Z_\ff 'Z_b '=Z_\ff Z_b=Z^2_\ff(g)$
}.
Nevertheless the two coupling constants exhibit a very peculiar
behaviour under the RG flow, moreover there are not acceptable fixed
point beside the trivial one ($\ll=\nu=0$).
In order to show this explicitly, let us compute the $\bb$ and $\gg$
functions, we get in $D=4-2\ee$:
\eqa
\bb_\ll&=&
      -2\ee\ll
      +{1\over (4\pi)^2}8(\ll^2-3\ll\nu)
      +{1\over (4\pi)^4}16(-5\ll^3+18\ll^2\nu-21\ll\nu^2)
\nonumber\\
\bb_\nu&=&
      -2\ee\nu
      -{1\over (4\pi)^2}8(\ll^2-\ll\nu+2\nu^2)
      +{1\over (4\pi)^4}16(6\ll^3-15\ll^2\nu+12\ll\nu^2-11\nu^3)
\nonumber\\
\bb_\rr&=&
      -{1\over (4\pi)^2}4\rr(\ll+\nu)
      +{1\over (4\pi)^4}8\rr(\ll^2-2\ll\nu-3\nu^2)
\nonumber\\
\gg_m&=&
      {1\over (4\pi)^2}4m^2(\ll-\nu)
      -{1\over (4\pi)^4}24m^2(\ll-\nu)^2
\ena
Integrating the $\bb$ differential equations at one loop we get easily
(integrating firstly $g(\mu)$, then $\ll(\mu)$ and finally getting
$\nu(\mu)$ as the difference of the previous two functions):
%\eqa
%\ll(\mu)&=&{ g_0\over2}
%         {1\over 1-{g_0\over \pi^2}\log({\mu\over\mu_0})}
%         {1\over 1-\left( 1-{g_0\over \pi^2}\log({\mu\over\mu_0})
%                        \right)^{1/2}
%                   \left( 1-{g_0\over 2\ll_0} \right) }
%\nonumber\\
%\nu(\mu)&=&{ g_0\over2}
%         {1\over 1-{g_0\over \pi^2}\log({\mu\over\mu_0})}
%\left(  {1\over2}
%         {1\over 1-\left( 1-{g_0\over \pi^2}\log({\mu\over\mu_0})
%                        \right)^{1/2}
%                   \left( 1-{(g_0\over 2\ll_0} \right) }
%     -1
%\right)
%\nonumber\\
%\ena
\eqa
\ll(\mu)&=&{ g_0\over2}
         {1\over 1-{g_0\over \pi^2}x}
         {1\over 1-\left( 1-{g_0\over \pi^2}x
                        \right)^{1/2}
                   \left( 1-{g_0\over 2\ll_0} \right) }
\nonumber\\
\nu(\mu)&=&{ g_0\over2}
         {1\over 1-{g_0\over \pi^2}x}
\left(  {1\over2}
         {1\over 1-\left( 1-{g_0\over \pi^2}x
                        \right)^{1/2}
                   \left( 1-{g_0\over 2\ll_0} \right) }
     -1
\right)
\nonumber\\
\ena
where $\ll_0=\ll(\mu_0),\nu_0=\nu(\mu_0)$ and $g_0=g(\mu_0)=\ll_0-
\nu_0$ and $x=log({\mu\over\mu_0})$.

It is easy to see that $\ll(\mu)$ has two singularities (fig. 5):
one is the usual Landau pole of $\ff^4$ at
$x_L=\log({\mu_L\over\mu_0})={\pi^2\over g_0}$
and the other is at
$$x_P=\log({\mu_P\over\mu_0})=
  {\pi^2\over g_0}
\left(1-{1\over
         {(1-{g_0\over 2\ll_0})^2} } \right)
$$
This latter singularity is shared also by $\nu(\mu)$ because it takes
place at a finite value of $g(\mu)$. What is the meaning of this singularity?
There are two possibilities; it could be
 either a breakdown of the perturbative expansion ( and in this case
it is probably related to the specific formulation of the theory)
or a problem intrinsic to the theory (\cite{BCCZ}).
What makes more reliable the first possibility is that this pole is
present even when the theory is free, i.e. setting $\ll=\nu$, in this
case $\bb_\ll=\bb_\nu=-{1\over\pi^2}\ll^2$ would lead to a singularity
in $x_P=\log({\mu_P\over\mu_0})=-{\pi^2\over \ll_0}$

\noindent
There is also an other singular point of the perturbative expansion of
the theory:
it happens when $\nu(\mu)$ crosses the zero and then it becomes
negative, in that case
the theory is not bounded from below anymore as can easily seen from
(\ref{prova}); this happens for

$$\log({\mu\over\mu_0})>x_Z={\pi^2\over g_0} \left(1-{1\over
{(2-{g_0\over\ll_0})^2} }\right) $$

The singular points of $\rr$ are at $x_L$ where it diverges and
at $x_P$ where it vanishes,
but differently from the previous singular behaviours, these
can be eliminated setting $\rr=0$, that, as shown by
(\ref{1loop},\ref{2loop}), does not change the physics.

\sect{Conclusion.}
In this paper we have demonstrated that the topological theory
we proposed is renormalizable and we have explicitly computed its
two loop perturbative expansion,
however the main aim of our approach, the exact computation
of the critical indexes of SAW, has revealed unreachable, nevertheless
this topological theory reveals interesting features:
\begin{enumerate}
\item even in the delta gauge it is not finite;
\item it has two phases, one of which has an explicit breaking of
      the topological character.
\end{enumerate}

There is an heuristic way to see immediately the existence of two
phases. It consists of a mean field approximation in which all the
fields are constants, whence the action can be written as
$$S=( \rr |b|^2+4\ll w^2-2i~m^2z+4\nu z^2)~ V$$
where $V$ is the volume, $w=Im (b^*\ff)+i~ Re(\xi^*\eta)$ and
$z=Re (b^*\ff)-i~ Im(\xi^*\eta)$. If we try to minimize
this action and
we consider that it is limited from below, we get immediately that
$b=w=0$ and $z=-{i m^2\over 4\nu}$, that implies
$$ <b^*\ff+\ff^* b>=0 \ne <\xi^*\eta-\eta^*\xi>={i m^2\over 2\nu}$$
while they should be equal in order not to break
the BRST symmetry.

\vskip 1cm
{\large \bf Acknowledgements.}

\noindent
I want to thank M. Caselle, D. Cassi, F. Gliozzi, R. Iengo and
N. Maggiore for useful discussions.

\end{document}